\documentclass[aps,pra,twocolumn]{revtex4-1}  

\usepackage{graphicx}
 
\begin{document}

\title{Optimal feedback control for the rapid preparation of a single qubit}

\author{Kurt Jacobs}

\affiliation{Centre for Quantum Computer Technology, Centre for Quantum
Dynamics, School of Science, Griffith University, Nathan 4111, Australia}
\affiliation{Present address: U.S. Army Research Laboratory, Computational and Information Sciences Directorate, Adelphi, Maryland 20783, USA}

\begin{abstract}
We consider the use of feedback control during a measurement to increase the 
rate at which a single qubit is purified, and more generally the rate at 
which near-pure states may be prepared. We derive the optimal bang-bang
algorithm for rapid state preparation from an initially completely mixed state
when the measurement basis is unrestricted, and evaluate its performance
numerically. We also consider briefly the case in which the measurement basis
is fixed with respect to the state to be prepared, and describe the qualitative
structure of the optimal bang-bang algorithm. 
\end{abstract}


\maketitle 

\section{INTRODUCTION}
\label{intro}  

In preparing a quantum system in a known pure state, one often starts with a
mixed state, either because noise processes have rendered the initial state
unknown, or because the system is entangled with other quantum systems. As a
result, state-preparation requires the use of both measurement and unitary
operations (Hamiltonian evolution). The measurement provides the required
purification, and the unitary evolution the ability to pick a particular final
state. 

However, it turns out that for continuous measurements there exists a nontrivial
interplay between the measurement process and unitary operations: the latter,
if applied during the measurement, change the rate at which the system is
purified~\cite{rapidP}. Thus, a process of {\em Hamiltonian feedback} (that is, 
unitary operations performed conditionally upon the continuous output of the
measurement process) plays the role not only of picking the appropriate final
state, but also of determining the speed of projection, and thus ultimately of
preparation. Note that while measurements are often treated as being
instantaneous, all real measurements are continuous in that they take some time
to act; measurements may be treated as instantaneous only if the measurement 
time-scale is much shorter than all other time-scales relevant to the given problem.

The problem we consider here is the preparation of a single qubit in the
fastest time. This is a particular instance of a problem in the domain of
quantum feedback
control~\cite{BelavkinBell,BelavkinLQG,Belavkin,DJ,DHJMT,Wisefb},  and like all
such problems it only makes sense when some or all of the resources at the
disposal of the controller are finite. Thus we will be interested in the
optimal speed of preparation given finite rates of measurement and/or
Hamiltonian evolution. We will find that even this simple task of preparing a
single qubit, employing the most straightforward measurement process, has a
non-trivial structure.

While the dynamics of closed quantum systems is linear, that of observed
quantum systems is often non-linear. The problem of feedback control of quantum
systems is therefore generally non-linear (even for very simple systems), and
our preparation problem is no exception. Few analytic results exist for
optimal control algorithms for non-linear systems. However, the tractability of
the problem depends not only upon the dynamics of the system, but also the
nature of the resource restrictions and how they are applied. 

One method of enforcing resource restrictions is simply to fix the maximal
resources allowed, and to optimize the desired behavior of the algorithm under
this restriction. In this case the functional to optimize is merely a function
of the system dynamics under the action of the feedback. An alternative
procedure is to make the resources flexible, and to minimize a `cost' function
which contains both a contribution from the amount of feedback required, and
the resulting system dynamics. In this case `optimality' means optimality with
respect to both the amount of resources used, and the resulting dynamics. The
relative weighting between the cost of control and the cost of undesired
dynamics is a free parameter. 

While the second of the above approaches is usually unsolvable for non-linear
systems, a solution is often clear for the first if the constraint is merely on
the strength of the applied feedback (the size of the applied force). That is,
when one assumes that the rate at which this force can be changed is
effectively unlimited on the relevant time scales of the system. The resulting
algorithms are such that at any particular time the applied force is set at its
extreme value in one direction or the other (for a one dimensional problem).
The resulting feedback algorithms are referred to as `bang-bang' control, a
term that presumably originates from the sound that a mechanical controller
makes when switching backwards and forwards between two extreme values.

One can view our problem as consisting of a set of control problems
corresponding to increasingly realistic resource constraints. We will find that
when the strength of the measurement is fixed and the Hamiltonian is unlimited,
the optimal algorithm may be found analytically, as well as its performance.
When we next impose a limitation on the strength of the feedback Hamiltonian
(but nothing else), there are broadly speaking two problems: one in which the
measurement basis is free to vary and the second in which it is fixed. For the 
former the optimal bang-bang algorithm is easily obtained, but its performance
must be calculated numerically. For the latter it is only possible to derive
the overall structure of the algorithm. Both the quantitative details and
performance must be calculated numerically.

In Section~\ref{rapidproj} we introduce continuous measurement and analyze
rapid projection via feedback. Following reference~\cite{rapidP} we derive the
optimal algorithm under the assumption of unlimited Hamiltonian resources, and
its performance. The latter may be regarded as an an upper bound on all
algorithms possible with a finite Hamiltonian. In Section~\ref{flex} we
consider rapid state preparation with finite Hamiltonian resources. In the
first part we consider the case in which the measurement basis is under our
control, derive the optimal algorithm and calculate numerically its perform as
a function of the magnitude of the Hamiltonian. This allows us to see how large
a Hamiltonian is required to approach the upper bound in the previous section.
In the second part we examine the case in which the measurement basis is fixed
(and corresponds to the preparation basis), and describe the structure of the
optimal bang-bang algorithm.

Before we begin we note that while very little has been written to date on 
optimizing the speed of state-preparation, recently there have appeared two
related works by van Handel, Stockton and Mabuchi~\cite{HSM,SHM} which consider
the deterministic preparation of one of the eigenstates of a continuously
measured observable by using feedback during the measurement. There are also a
number of articles which are concerned more generally with the feedback control
of two-state quantum systems~\cite{Korotkov,RK,DJJ,WMW,BEB}.

\section{Rapid Purification via Feedback} 
\label{rapidproj}

All physical measurements are continuous - that is, every measurement extracts information at a finite rate. When this rate is much larger than all other relevant time-scales, then one can approximate the measurement by a von Neumann measurement. Otherwise one must describe the continuous extraction of information explicitly. To describe a continuous measurement classically, one simply writes down the relationship of the measurement results, $r(t)$ (which are a function of time), to the true value of the measured quantity, $y(t)$, which is also in general a function of time. For virtually all applications in which the stream of output results is a continuous function of time, this relationship is
\begin{equation}
   dr = y(t) dt + c dV ,
\end{equation}
where $c$ is some constant, and $dV$ is an increment of the Gaussian noise process referred to as the Wiener process. (For readers not familiar with the Wiener process, we note that easily accessible treatments may be found in references~\cite{WienerIntroPaper} and~\cite{KJPhD}.) Gaussian noise is appropriate for virtually all continuous measurements in which $r(t)$ a continuous function because of the central limit theorem. It turns out that the measurement record $r$ may also be written in terms of the expectation value of $y$ at each time. Specifically 
\begin{equation}
   dr = \langle y(t) \rangle dt + c dW ,
   \label{rec}
\end{equation}
where $dW$ is another zero-mean Wiener process (Gaussian white noise), uncorrelated with $dV$. 

One then uses Bayes' theorem~\cite{Jaynes,BayesIntro} to obtain an equation which gives the evolution of the observer's state-of-knowledge regarding $y(t)$ (that is, her probability density for $y(t)$) as the measurement progresses. This equation is called the Kushner-Stratonovich equation~\cite{KSeq}, and for a $y$ which does not change with time, is given by 
\begin{equation}
  dP(y,t) = 2\sqrt{2k}(y - \langle y \rangle) P dW ,
  \label{KS}
\end{equation} 
where we have chosen $k = 1/(8c^2)$ to simplify the treatment in what follows. The quantity $k$ characterizes the rate at which the measurement extracts information about $y$, and we will refer to it as the {\em measurement strength}.

Quantum mechanically, one can describe a continuous measurement by using a sequence of ``weak'' measurements, where one takes the continuum limit of an infinite number of infinitely weak measurements~\cite{CMnotes}. In doing so, one obtains the precise quantum equivalent of the classical continuous measurement given by Eqs. (\ref{rec}) and (\ref{KS}). In the quantum case the observer's state-of-knowledge is given by the density matrix. If the measurement extracts information about an observable $\hat{y}$, then the quantum equivalent of the Kushner-Stratonovich equation is
\begin{equation}
  d\rho = -k[\hat{y},[\hat{y},\rho]] dt + \sqrt{2k}(\hat{y}\rho + \rho\hat{y} 
                                        - 2\langle \hat{y} \rangle \rho ) dW .
\label{genSME}
\end{equation} 
This equation is referred to as a {\em Stochastic Master Equation} (SME).
If $\hat{y}$ commutes with the initial density matrix, then this quantum equation reduces to the classical equation (Eq.(\ref{KS})), because quantum measurement theory reduces to classical measurement theory (as it must) when all relevant quantities commute~\cite{sk}. In this case $P(y,t)$ is simply the diagonal of $\rho(t)$.

Let us consider now a continuous measurement of the $z$-component of spin of a spin-1/2 particle. The SME for this measurement is given by setting $\hat{y} = \sigma_z$ in Eq.(\ref{genSME}) above. If we write the density matrix in terms of the Bloch vector, ${\bf a} = (a_x,a_y,a_z)$, as
\begin{equation}
  \rho(t) = \frac{1}{2} (I + {\bf a}\cdot{\bf \sigma}) ,
\end{equation}
then the SME becomes
\begin{eqnarray}
  d a_x & = & -(4k dt + a_z \sqrt{8k} dW) a_x , \\
  d a_y & = & -(4k dt + a_z \sqrt{8k} dW) a_y , \\
  d a_z & = & (1 - a^2_z) \sqrt{8k} dW .
\end{eqnarray}
From these it is easy to show that the relation between $a_x$ and $a_y$ is a constant of the motion, and thus the initial angle of the Bloch vector in the $x$-$y$ plane, $\phi = \arctan(a_x/a_y)$, remains unchanged throughout the measurement. We can thus reduce the equations of motion to two variables. If we define $\Delta = \sqrt{a_x^2 + a_y^2}$, being the length of the projection of the Bloch vector in the $x$-$y$ plane, then the equations of motion become
\begin{eqnarray}
  d \Delta & = & -(4k dt + a_z \sqrt{8k} dW) \Delta , \label{SMED} \\
  d a_z & = & (1 - a^2_z) \sqrt{8k} dW .
  \label{SMEaz}
\end{eqnarray}
Due to symmetry the angle $\phi$ plays no role in the dynamics. 

The purity of the density matrix, $\mbox{Tr}[\rho^2]$, being the squared length of the Bloch vector, characterizes the observers certainty regarding which pure state the system is in. If the observer has no information regarding the state of the system, this corresponds to a uniform distribution over all pure states (or, alternatively, equal probabilities for the two eigenstates of $\sigma_z$). In this case the density matrix is described as being ``completely mixed'', and is proportional to the identity matrix. In this case the purity is minimal. If, on the other hand, the observer knows the system to be in a specific pure state, then the purity obtains its maximal value of unity. To measure the observers uncertainty we will use $\bar{p} = 1 - \mbox{Tr}[\rho]$, and refer to this as the {\em impurity} (the ``bar'' is intended to represent negation). This quantity is often referred to as the {\em linear entropy}. In terms of the dynamical variables introduced above, the impurity is
\begin{equation}
  \bar{p} = 1 - \mbox{Tr}[\rho^2] = \frac{1}{2} ( 1 - \Delta^2 - a_z^2 ) .
\end{equation}

If the state of the system is initially completely mixed (completely uncertain), then $a_z(0) = \Delta(0) = 0$. While in this case $\Delta$ remains zero throughout the measurement, $a_z(t)$ is stochastically driven to $\pm 1$. That is, the state is projected onto one of the eigenstates of $\sigma_z$, which is the result of the measurement, and in this case each of these outcomes is selected with equal probability. In this case the measurement is also purely classical, since $\Delta$ is always zero, which means that the density matrix, $\rho$, commutes with $\sigma_z$ at all times. 

The stochastic equations which describe the measurement process (Eqs.(\ref{SMEaz}) and (\ref{SMED})) are non-linear. However, it is nevertheless possible to solve these equations by using an equivalent formulation (referred to as a ``linear quantum trajectory'') in which they become linear~\cite{JK} (see also~\cite{GG,WisemanLinQ}). The solution is 
\begin{widetext}
\begin{eqnarray}
  \Delta(t) & = & \frac{ \mbox{cosh}^2(\sqrt{2k}W(t)) - a_z(0) \mbox{sinh}^2(\sqrt{2k}W(t)) }{ \cosh(\sqrt{8k}W(t)) + a_z(0) \sinh(\sqrt{8k}W(t))} \Delta(0)  , \\
  a_z(t)    & = & \frac{ a_z(0) \cosh(\sqrt{8k}W(t)) + \sinh(\sqrt{8k}W(t))}{ \cosh(\sqrt{8k}W(t)) + a_z(0) \sinh(\sqrt{8k}W(t))} ,
\end{eqnarray}
\end{widetext}
and when $a_z(0) = 0$, this reduces to 
\begin{eqnarray}
  \Delta(t) & = & \frac{ \mbox{cosh}^2(\sqrt{2k}W(t))}{ \cosh(\sqrt{8k}W(t))} \Delta(0)  , \\
  a_z(t)    & = & \tanh(\sqrt{8k}W(t)). 
\end{eqnarray}
Here $W(t)$ is a random variable whose probability density is
\begin{eqnarray}
  && \!\!\!\!\!\! P(W,t) = \frac{e^{-4 k t}}{\sqrt{2\pi t}} \cosh(\sqrt{8k}W) e^{-W^2/(2t)} \\
  && \!\!\!\!\!\! = \frac{1}{\sqrt{4\pi t}} \left( e^{-(W + \sqrt{8k}t)^2/(2t)} + 
                                              e^{-(W - \sqrt{8k}t)^2/(2t)} \right) .
\label{Pdistclass}
\end{eqnarray}
Using this solution we can obtain an expression for the average impurity as the measurement proceeds. This is
\begin{equation}
  \bar{p}(t) = \frac{e^{-4k t}}{\sqrt{8\pi t}} \int_{-\infty}^{+\infty} \frac{e^{-x^2/(2t)}}{\cosh(\sqrt{8k}x)} dx .
  \label{Pclass}
\end{equation}
While this integral must be solved numerically, certain features may nevertheless be extracted from this expression for $\bar{p}$. One finds that initially $\bar{p}$ decays as an exponential at rate $4k$, and an approximate analytic expression may be derived in the long-time limit, to which we will return later. In addition, general properties of measurement make it clear that $\bar{p}$ decays monotonically with time~\cite{Nielsen,FJ}.

\begin{figure}
   \begin{center}
   \begin{tabular}{c}
   \includegraphics[width=8cm]{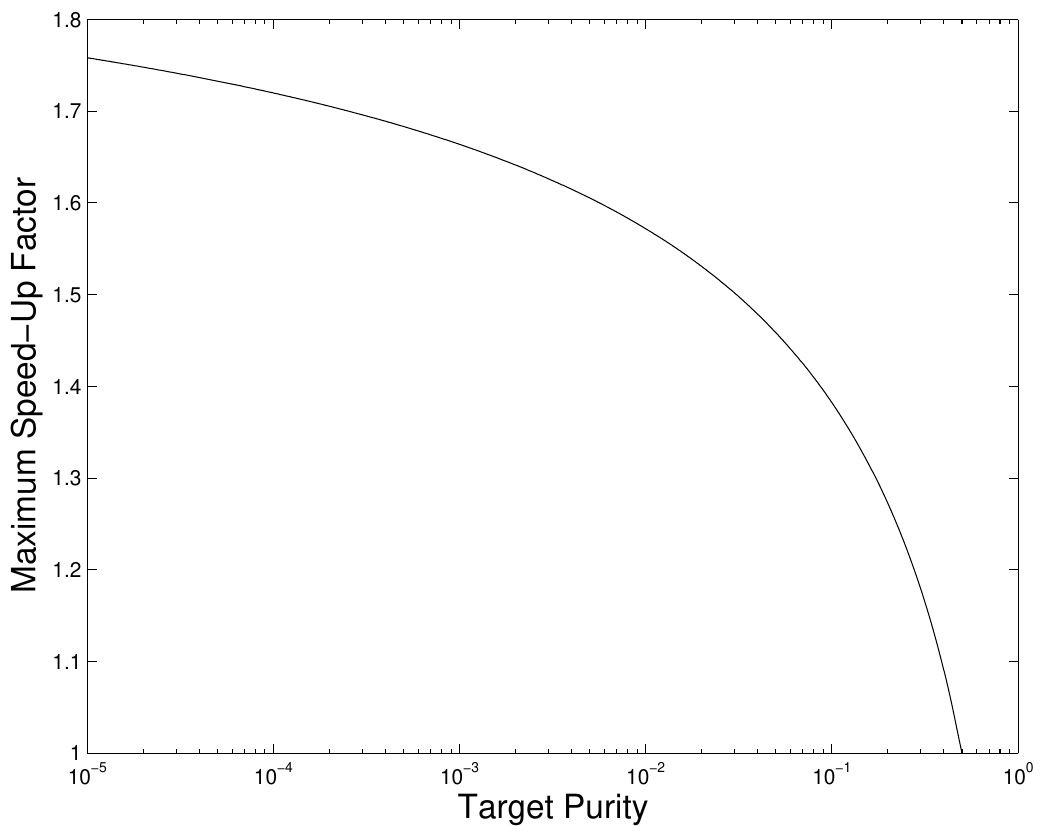}
   \end{tabular}
   \end{center}
   \caption[example] { \label{fig1}  The maximum possible average speedup in 
the time required to achieve a final target purity from an initially completely 
mixed state, for a single qubit, using Hamiltonian feedback. This bound on 
speedup is obtainable in the limit of a large feedback Hamiltonian. }
\end{figure} 

If this was a measurement on a classical bit, then the only way to increase the rate of reduction of the impurity would be to increase $k$, the strength of the measurement. However, quantum mechanically it is possible to increase this rate by using unitary operations during the measurement. To see this we must return to Eqs.(\ref{SMED}) and (\ref{SMEaz}), and calculate the rate of change of the square of the length of the Bloch vector, which is
\begin{eqnarray}
  d|{\bf a}|^2 & = & d\Delta^2 + da_z^2 = (1-a_z^2) (\Delta^2 - (1 + a_z^2)) 8k dt \nonumber \\ 
               &   & + a_z[(1 - a_z^2) - \Delta^2] \sqrt{8k} dW 
  \label{Eqmaga}
\end{eqnarray}
It is clear from this equation that for a given value of $|{\bf a}|^2$ the average rate of increase is greatest when $a_z = 0$ (that is, when the Bloch vector lies in the $x$-$y$ plane, perpendicular to the basis of the measurement). Furthermore, in this case the stochastic term vanishes, so that the increase in $|{\bf a}|^2$ is deterministic.

Since the rate of purification depends on the orientation of the Bloch vector one can apply unitary operations to affect this rate (unitary operations can be used to rotate the Bloch vector, but cannot change the length). The unitary operator
\begin{eqnarray}
U(\alpha) = \exp[-i(\alpha/2)(\cos(\phi),-\sin(\phi),0)\cdot{\bf \sigma}] ,
\end{eqnarray}
will rotate the Bloch vector by an angle $\alpha$ towards or away from the $x$-$y$ plane, while maintaining the angle $\phi$. The Hamiltonian which generates this motion is thus $H = \hbar\mu(\cos(\phi)\sigma_x -\sin(\phi)\sigma_y)$, and the angle of rotation obtained is $\alpha = 2\mu t$, where $t$ is the duration over which the Hamiltonian is in effect. The equations of motion describing such a rotation are simply
\begin{eqnarray}
  d\Delta & = & \mu a_z dt , \\
  d a_z & = & -\mu \Delta dt .
  \label{Hterms}
\end{eqnarray}

Now, we have seen from Eq.(\ref{Eqmaga}) that the rate of purification is maximal when the Bloch vector lies in the $x$-$y$ plane. However, from Eqs.(\ref{SMED}) and (\ref{SMEaz}), it is clear that even if the Bloch vector lies in this plane at any given time, it will not remain there; the noise term in the equation of motion for $a_z$ will kick it out. (Note that if $a_z=0$, then the only stochastic term that remains is that in the equation for $d a_z$.) To maintain the most rapid average rate of purification we must therefore continually rotate the Bloch vector back to the $x$-$y$ plane as the measurement proceeds. If we allow the Hamiltonian to be arbitrarily large, then we can choose the rotation terms (Eqs(\ref{Hterms})) to exactly cancel the noise term which kicks the Bloch vector out of the $x$-$y$ plane. (The Hamiltonian must be unbounded to achieve this, because $dW$ scales as $\sqrt{dt}$). Note that such a procedure is a feedback process, because the choice we make for the Hamiltonian at each point in time depends on the measurement result obtained at that time. In particular, to ensure that the Hamiltonian motion cancels the stochastic evolution due to the measurement, we must set
\begin{eqnarray}
  \mu(t) = \frac{\sqrt{8k}}{\Delta} \frac{dW}{dt} . 
  \label{Hfb}
\end{eqnarray}

There are a number of ways of calculating the evolution of the impurity under this feedback algorithm, but the most pedestrian turns out to be a little tricky; since the Hamiltonian term is now proportional to $dW$, we must consider the action of the Hamiltonian to second order. To perform the calculation we consider a single step of the measurement process, followed by a single step of the Hamiltonian feedback. At the start we set $a_z = 0$, and at the end of the two steps $a_z$ is once again zero. The first step for $\Delta$ is given by Eq.(\ref{SMED}), with $a_z=0$, and adding the second step we have 
\begin{eqnarray}
  d\Delta = -4k \Delta dt + \mu a_z dt - \frac{1}{2} \mu^2 \Delta (dt)^2 ,
\end{eqnarray}
where now $a_z = \sqrt{8k}dW$, being the $a_z$ which results from the first step, $\mu$ is given by Eq.(\ref{Hfb}) above, and we have included the action of the Hamiltonian to {\em second order}. Substituting in the relevant expressions, the result is 
\begin{eqnarray}
  d\Delta = -4k dt (\Delta - \frac{1}{\Delta}) ,
\end{eqnarray}
which is a purely deterministic differential equation. This might look a little odd - but if we write the equation for the impurity, $\bar{p} = 1/2 - 1/2 \Delta^2$, the behavior is immediately clear:
\begin{eqnarray}
  d\bar{p} = -8k \bar{p} dt ,
  \label{impopt}
\end{eqnarray}
which is simply an exponential decay at rate $8k$. 

This result seems quite remarkable - not only does the Hamiltonian feedback increase the rate at which the system is projected, but greatly simplifies it. While the classical result (given by Eq.(\ref{Pclass})) cannot be written in a fully analytic form, the quantum result with feedback is a simple exponential, and thus easily characterized by the rate $8k$. 

The feedback algorithm we have just presented is, in fact, optimal. The rate of increase of purity given by Eq.(\ref{impopt}) is the best that can be achieved for any feedback over any time period. While this is fairly clear from the above discussion, a systematic proof is given in reference~\cite{rapidP}. 

So how much does the optimal feedback algorithm speed up the purification process? While we do not have an analytic form for the `raw' purification as a function of time, we can obtain an expression valid in the limit as $t \rightarrow \infty$. To do so we need merely note that the integral in the expression for the raw impurity (Eq.(\ref{Pclass})) is independent of $t$ for large $t$. Thus for large $t$
\begin{equation}
  \bar{p}(t) = \frac{e^{-4k t}}{\sqrt{8\pi t}C} ,
  \label{PclassbigT}
\end{equation}
for some constant $C$.
If we set some very small impurity as the required target, then we can ask what is the ratio between the average classical time to reach this target, $t_{\mbox{\scriptsize c}}$, and the optimal quantum time, $t_{\mbox{\scriptsize opt}}$. When the target purity is sufficiently small (so that $t_{\mbox{\scriptsize c}}$ is sufficiently large), we have
\begin{equation}
  \bar{p}_{\mbox{\scriptsize target}} = \frac{e^{-4k t_{\mbox{\scriptsize c}}}}{\sqrt{8\pi t_{\mbox{\scriptsize c}}}C}  = e^{-8k t_{\mbox{\tiny opt}}},
\end{equation}
or
\begin{equation}
  \frac{t_{\mbox{\scriptsize opt}}}{t_{\mbox{\scriptsize c}}} = \frac{1}{2} + 
              \frac{\ln t_{\mbox{\scriptsize c}}}{8k t_{\mbox{\scriptsize c}}} + \frac{\ln  8\pi C}{8k t_{\mbox{\scriptsize c}}} .
\end{equation}
This tends to a value of $1/2$ as $t_{\mbox{\scriptsize c}} \rightarrow \infty$, so that in the long-time limit the time taken to reach the target impurity is reduced by a factor of 2. A numerical solution of the integral expression for the raw $\bar{p}$ confirms that the speed-up factor increases with time monotonically, and thus the maximal speedup factor is 2. The speedup factor as a function of the target impurity is plotted in Figure~\ref{fig1}.

While the optimal feedback algorithm presented above allows us to find the theoretical maximum extent to which unitary operations can increase the rate of entropy reduction rendered by a continuous measurement, the algorithm assumes unlimited Hamiltonian resources. We have not, therefore, determined how large a Hamiltonian one would require in order to get close to the theoretical limit, and what speedup is achievable with a given strength of Hamiltonian. We have also been concerned merely with the purification of a quantum system, rather than the preparation of a particular pure quantum state. In the next section we will consider these questions, and the implications of the above results for rapid state-preparation.

\section{Rapid State-Preparation with Finite Hamiltonian resources} 
\label{stateprep}

In the light of the above analysis, how can one prepare a given pure state in the fastest time? If the Hamiltonian resources at ones disposal are sufficiently large as to be effectively unlimited, then the answer is clear: one uses the Hamiltonian to keep the Bloch vector unbiased with respect to the measurement basis while the measurement proceeds, and then as soon as the desired level or purity has been achieved, one rotates the system to the desired state. 
However, what happens if the Hamiltonian resources are finite? To place a minimal restriction on the Hamiltonian we can put an upper bound on trace of its square:
\begin{equation}
  \mbox{Tr}[H^2] \leq \lambda^2 ,
\end{equation}
for some constant $\lambda$.
This restriction is unitarily invariant, so the controller is able to perform any rotation on the Bloch sphere, being limited only in the {\em rate} of this rotation. Under this resource limitation, one can identify two situations, one in which the basis of the measurement is free to move, and one it which it is fixed. We now consider each of these situations in turn.

\begin{figure}
   \begin{center}
   \begin{tabular}{c}
   \includegraphics[width=8cm]{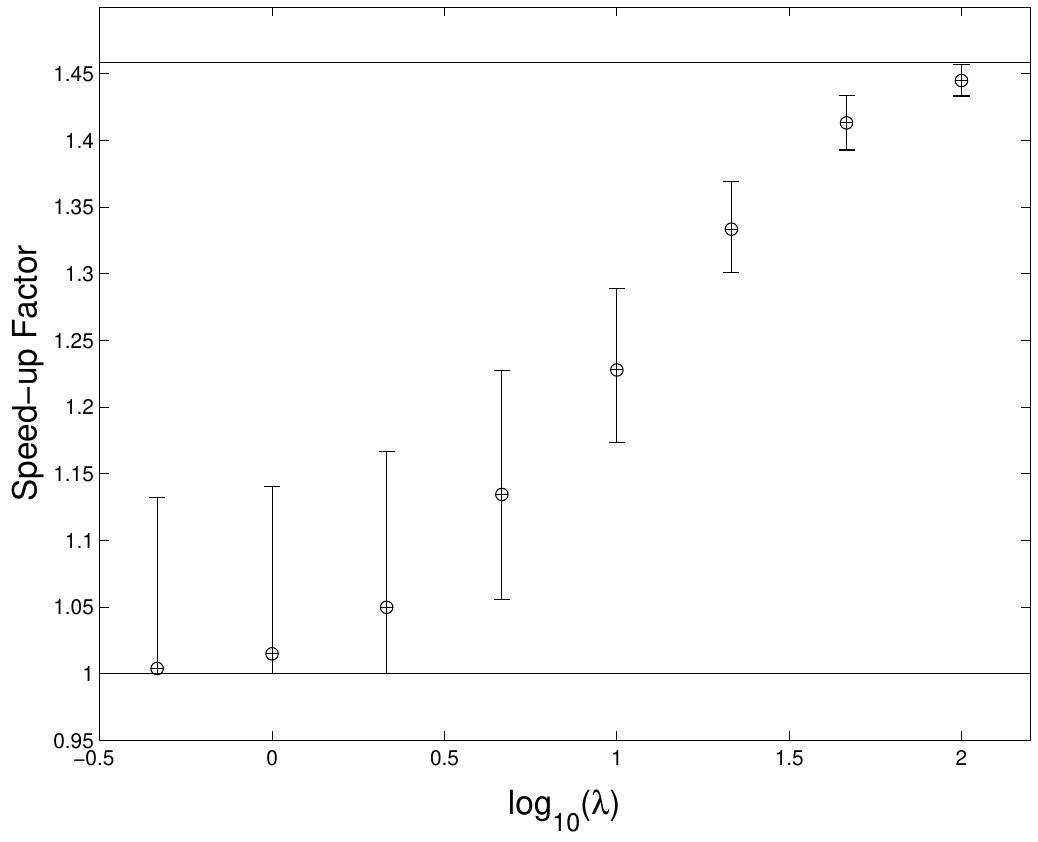}
   \end{tabular}
   \end{center}
   \caption[example] { \label{fig2} 
The average speed-up in the time to obtain a purity of $0.95$, from an
initial purity of $5\times 10^{-4}$, for different values of the Hamiltonian feedback
strength, $\lambda = \mbox{Tr}[H^2].$ The upper horizontal line is the maximal
speed-up obtainable in the limit of large a large feedback, and the lower line
is unity (no speed-up). Each plotted point was obtained by averaging over 4000
realizations. Note that the plotted points do not fluctuate with respect to one
another as much as the error bars would suggest - this is because the same
random number sequence has been chosen in generating each point, with the
result that the errors are highly correlated across the points.}
\end{figure} 

\subsection{Flexible measurement}
\label{flex}

If we wish to prepare a state $|\psi\rangle$ from an initially completely mixed state, and we are free to chose the basis in which our measurement is performed, then we can choose this basis to be unbiased with respect to the state $|\psi\rangle$. Thus is equivalent to being able to choose our target state $|\psi\rangle$ as lying on the $x$-$y$ plane, when we measure in the $\sigma_z$ eigenbasis. 

The most rapid path to the state $|\psi\rangle$ is the path which lies in the $x$-$y$ plane, along the line on which the states have the same $\phi$ as the target $|\psi\rangle$. (That is, the line which connects the center of the Bloch ball with the state $|\psi\rangle$.) This is clear because states in the $x$-$y$ plane give the fastest rate of purification, and thus if $\phi$ is chosen correctly at the start (we will return to this point below), no further rotation is required at the end of the purification process. 

Since we have finite Hamiltonian resources, the feedback algorithm will not be able to ensure that the state of the system remains exactly unbiased with respect to the 
measurement. However, since the average rate of purification decreases with increasing distance from the $x$-$y$ plane, then, so long as $\phi$ can be selected correctly at the start, the optimal algorithm is simply the one which keeps the state closest to this plane. Since we have not imposed restrictions on the rate at which our feedback can respond, the optimal algorithm is thus a bang-bang algorithm which rotates the state up or down towards the $x$-$y$ plane with maximal speed, depending on whether the state is above or below the plane. The remaining issue is the initial selection of the angle $\phi$. At the start of the process, since we are assuming that the initial state is completely mixed, the initial action of the measurement is to point the state along the $z$-direction. Since this is the case, in choosing an initial Hamiltonian to rotate the state towards the $x$-$y$ plane, we are also free to choose the initial $\phi$.

The optimal bang-bang algorithm for preparing a given nearly-pure state from an
initially completely unknown state is thus clear. The dynamical equations,
including this feedback, are now 
\begin{eqnarray} 
  d \Delta & = & -(4k dt + a_z\sqrt{8k} dW) \Delta + \mu(t) a_z dt , \\ 
  d a_z & = & (1 - a^2_z) \sqrt{8k} dW - \mu(t) \Delta dt. 
  \label{SMEflex} 
\end{eqnarray} 
where  
\begin{eqnarray}
  \mu(t) & = & \left\{ \begin{array}{r}  \lambda \;\; , \;\;\; a_z > 0 \\
                                        -\lambda \;\; , \;\;\; a_z < 0   
	               \end{array} \right. . 
\end{eqnarray} 
However, in contrast to the algorithm in the previous section, the resulting
dynamics cannot be solved exactly; not even the technique of ``linear quantum
trajectories''~\cite{JK,WisemanLinQ} is sufficient. One must therefore solve
these equations numerically. We do this and present the speed-up achieved for
one value of the target purity as a function of feedback strength $\lambda$ in
Figure~\ref{fig2}. This shows us that a feedback strength which is 100 times
the measurement strength $k$ is sufficient to get within $5\% 
$ of the theoretical limit. 

We have examined the case in which the initial state of the qubit is completely
mixed. The resulting optimal bang-bang algorithm is also applicable to the case
in which the initial state is not completely mixed, but is a mixture of the
identity and the state to be prepared -- that is, lies along the line in the
Bloch ball joining the target state  to the center. However, this algorithm is
no longer applicable when the initial state does not lie on this line. Further,
it is not clear that the optimal algorithm in such a case can be obtained
analytically. While we will not consider this problem in detail here, we
describe briefly the source of the complexity. When the initial state has the
wrong $\phi$, Hamiltonian resources are required to change the angle of the
state in the $x$-$y$ plane (that is, to change $\phi$). If the controller
chooses to rotate the state in the $x$-$y$ plane during the measurement, then
this reduces the ability of the controller to keep the state {\em in} the
$x$-$y$ plane, and purification will be slower. However, if the controller
waits until the desired purification has been achieved, and then  rotates the
state, this takes extra time at the end of the process. It is thus not clear
how best the controller should apportion resources between purification and
orientation during the measurement, and this remains an open problem for future
work.

\subsection{Fixed measurement}
\label{fixed}

In the preceding section we considered rapid state preparation when the
observer was able to freely choose the basis in which the measurement was made
(i.e. the observable being measured). In this case we were able to obtain a
simple optimal bang-bang algorithm for rapid state preparation when the initial
state was completely mixed. In this section we consider the situation in which
the observable is fixed with respect to the target state. In this case we will
not be able to obtain the precise optimal bang-bang algorithm analytically, but
we will be able to sketch the qualitative structure.  

If the measurement basis is fixed, then we cannot choose this basis to be
unbiased with respect to the target state. This means that the path of most
rapid purification will not include the target. Let us assume for concreteness
that the measurement basis is the $\sigma_z$ basis, and that the available
Hamiltonian resources are such that the time required to rotate a state in the
$x$-$y$ plane to the target state is less than that required to obtain the
target purity. If the controller decides to keep the state in the $x$-$y$ plane
until the required purity is reached, and then rotates the state to the target,
the time required is the optimal time to purify, plus the final rotation time.
It is clear that one can do better. This is because, the purification rate off
the $x$-$y$ plane, and thus closer to the target state, is not zero. Thus, if
the controller starts to rotate the state towards the target shortly before
obtaining the final purity, some time will be lost, but less than waiting until
the end to perform the rotation. In fact, so long as the time taken to purify
during the rotation is not greater than that required for the rotation, it is
better to rotate before reaching the required purity.

Conversely, a similar argument tells us that it is best not to rotate the state
out of the $x$-$y$ plane and towards the target if the average time to
purification is greater than that required for the rotation. This is because,
during any excess time required at the end for the remaining purification, this
purification will happen at a non-optimal rate. We can therefore conclude that
there exists an optimal point of purity (distance from the center of the Bloch
ball) at which it is no-longer desirable to keep the state on the $x$-$y$
plane. Further, the same reasoning holds at any angle off the $x$-$y$ plane
towards the target state, but as the angle gets closer to the target, the time
required to rotate to the target is less. As a result, as the angle to the
target decreases, the point of purity at which one should rotate towards the
target is ever larger (ever closer to the edge of the Bloch ball), ultimately
reaching the target purity at the target state.

There exists therefore a critical surface, inside which one should rotate the
state at the maximum available rate so as to keep it close to the the $x$-$y$
plane. Outside this surface one should rotate the state as fast as possible
towards the target state. That this rotation should be done as fast as possible
follows from the fact that one wants to spend as little time as possible in the
region off the $x$-$y$ plane; one thus waits for as long as possible before
rotating towards the target, and thus must rotate with maximum speed when the
time comes.

\begin{figure}
   \begin{center}
   \begin{tabular}{c}
   \includegraphics[width=8cm]{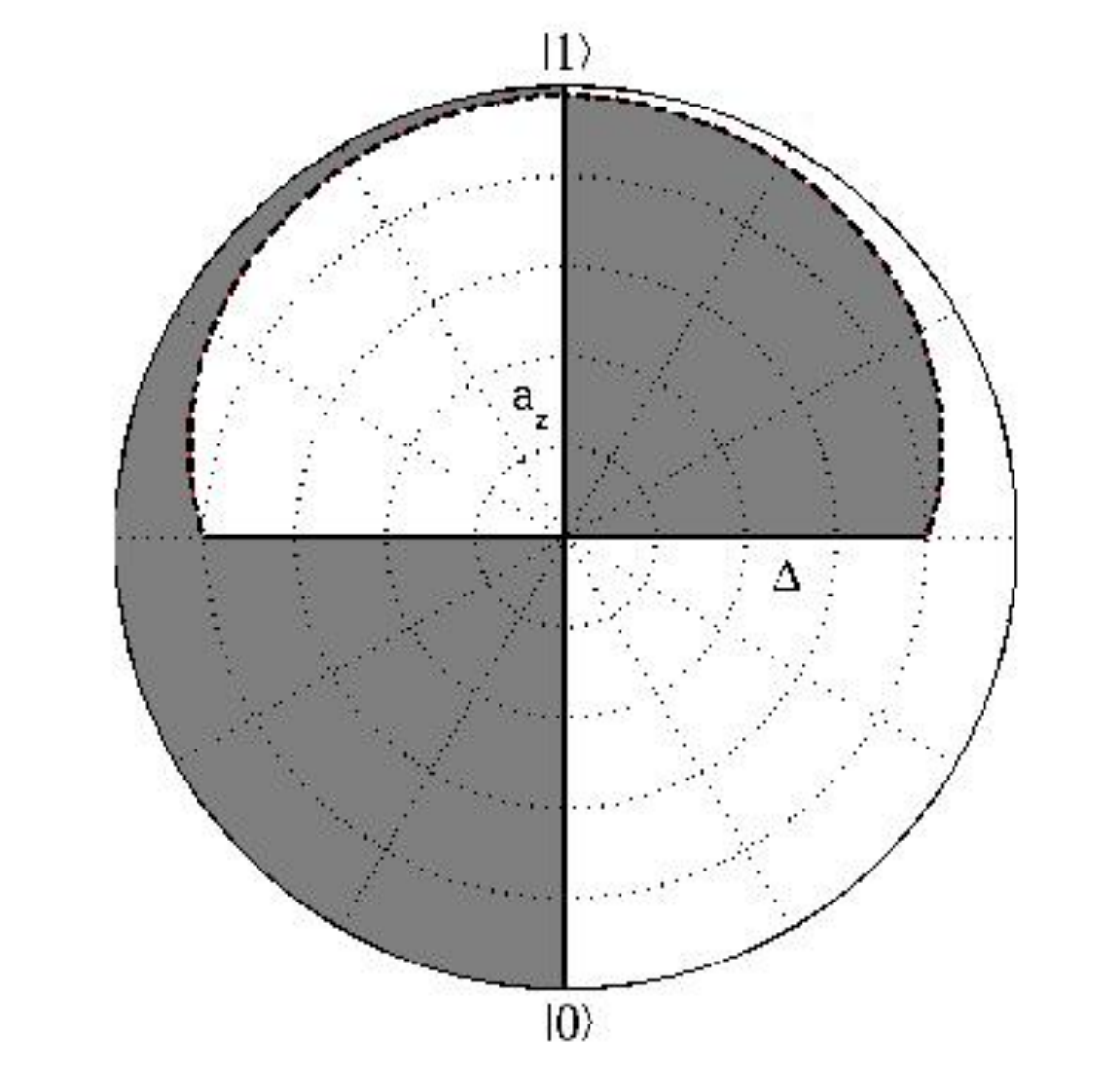}
   \end{tabular}
   \end{center}
   \caption[example] { \label{fig3} A diagram giving the qualitative structure of the optimal bang-bang feedback algorithm for the rapid preparation of a nearly-pure state close to $|1\rangle$, when the measurement is fixed in the $z$-basis $\{|0\rangle,|1\rangle\}$. The grey portions denote rotation in the anti-clockwise direction at the maximum rate, and the white regions denote clockwise rotation. The precise location of the dotted curves would have to be obtained numerically.}
\end{figure} 

The qualitative form of the optimal bang-bang algorithm which results from the
above reasoning is displayed in Figure~\ref{fig3} (for the case in which the
measurement is in the $\sigma_z$ basis, and the target state is $|1\rangle$).
The algorithm involves choosing the Hamiltonian so as to rotate the state
towards the $x$-$y$ plane ($a_z=0$), until a critical line is reached (the
dotted lines in Figure~\ref{fig3} - there are two due to symmetry). These lines
start at the $x$-$y$ plane, and become closer to the edge of the Bloch ball as
they move towards the target state, reaching the target purity at that point.
Once the state of the system crosses these lines, the Hamiltonian is reversed
so as to rotate towards the target.

How might one go about calculating the exact form for the critical lines in
Figure~\ref{fig3}? The time taken to reach the target is ultimately a
functional of the critical line. One could therefore simulate the feedback
algorithm to obtain the average time-to-target, and seek to optimize this by
varying the critical line. In implementing such a procedure one would ideally
start with a guess for the critical curve based on approximate arguments. Even
so, one would expect this method to require considerable numerical resources. 

The position of the critical line is determined by the average time required to
reach the target purity from any given point. However, this average time is
itself a function of the feedback algorithm, and thus the positioning of the
critical line - wherein lies the difficulty. It might be possible to obtain an
approximate solution using a numerical algorithm which discretizes the angle to
the target, and considers small angles first. That is, calculates the remaining
purification time at a small angle from the target, assuming that feedback at
all larger angles involves rotating towards the target, and then once this is
determined works backwards to larger angles. However, it is not clear that such
an algorithm would reduce the numerical resource requirements. In any case, the
task of obtaining the precise form of the optimal feedback algorithm remains an
open problem for future work. 

\vspace{7mm}

\section{Conclusion} 

We have seen that the rate at which a quantum state is purified by a continuous
measurement is not isotropic on the Bloch sphere; it depends on the angle
between the state and the measurement basis. This curious fact means that the
task of rapid state preparation using feedback control is non-trivial. It is
possible to derive the optimal bang-bang algorithm for purification in a
special case in which one is free to choose the measurement basis, but it is
not clear that a fully analytic solution will exist for the general case.
Nevertheless, it is possible to describe the qualitative structure of the
algorithm for the general case, and this algorithm does not involve taking the
shortest path to the target state. 

\vspace{-3mm}

\acknowledgments       
 
This work was supported by the Australian Research Council and the State of 
Queensland. Part of this research was performed using the supercomputing 
resources located at the Queensland Parallel Supercomputing Foundation. 


\end{document}